\documentstyle{elsart}

\def\openone{\leavevmode\hbox{\small1\kern-3.3pt\normalsize1}}

\begin{document}

\begin{frontmatter}

\title{Universal spectral correlations of the Dirac operator at finite
  temperature}
\author[Heidelberg]{Thomas Guhr} and 
\author[Heidelberg,Muenchen]{Tilo Wettig}
\address[Heidelberg]{Max-Planck-Institut f\"ur Kernphysik, 
         Postfach 103980, D-69029 Heidelberg, Germany}
\address[Muenchen]{Institut f\"ur Theoretische Physik,
  Technische Universit\"at M\"un\-chen, D-85747 Garching, Germany}

\date{\today}

\begin{abstract}
  Using the graded eigenvalue method and a recently computed extension
  of the Itzykson-Zuber integral to complex matrices, we compute the
  $k$-point spectral correlation functions of the Dirac operator in a
  chiral random matrix model with a deterministic diagonal matrix
  added.  We obtain results both on the scale of the mean level
  spacing and on the microscopic scale.  We find that the microscopic
  spectral correlations have the same functional form as at zero
  temperature, provided that the microscopic variables are rescaled by
  the temperature-dependent chiral condensate.\\[4mm]
\end{abstract}

\end{frontmatter}

\section{Introduction}
\label{sec0}

In the past few years, random matrix models have started to be used in
order to understand details of the spectral properties of the Dirac
operator of non-abelian gauge theories~\cite{Nowa89,Simo91,Shur93,%
Verb93,Verb94b,Naga93,Verb94a,Verb94c,Verb94d,Hala95a,Smil95,Hala95b,%
Verb96a,Jack96a,Wett96,Step96a,Step96b,Nowa96,Jack96b,Jack96c,Verb96b}.
Since random matrix models are free of dynamical input, one expects
them to describe universal spectral properties, i.e., properties which
do not depend on the specific details of the dynamics.  In the case of
non-relativistic quantum mechanics, it has long been known for a
variety of physical systems that random matrix models are able to
describe the universal spectral correlations on the scale of the mean
level spacing~\cite{Bohi84,Meht91,Haak91}. In the case of the Dirac
operator of QCD the chiral structure has to be incorporated, requiring
the introduction of the chiral random matrix ensembles~\cite{Shur93}.
The symmetry constraints lead to the definition of the chiral Gaussian
Orthogonal (chGOE), Unitary (chGUE) and Symplectic (chGSE) Ensembles.
A precise classification can be found in Ref.~\cite{Verb94a}.  Our
chiral random matrix model is based on unitary symmetry and will be
defined in Sec.~\ref{sec1}.

For the case of QCD with two colors, the universality of the spectral
correlations in the bulk of the spectrum of the lattice Dirac operator
was demonstrated convincingly in Ref.~\cite{Hala95a} where predictions
from random matrix theory were compared with lattice data obtained by
Kalkreuter.  However, the fact that the chiral symmetry of the Dirac
operator is spontaneously broken implies that another type of
universality exists concerning the microscopic limit of the spectral
density.  The spectral density is defined by
\begin{equation}
  \label{eq0.2}
  \rho(\lambda)=\langle\sum_n \delta(\lambda-\lambda_n)\rangle_A \:,
\end{equation}
where the $\lambda_n$ are the eigenvalues of the Dirac operator
$iD=i\gamma_\mu\partial_\mu+\gamma_\mu A_\mu$ (in euclidean space) and
the average is over all gauge field configurations $A$ weighted by the
euclidean QCD action.  Note that $iD$ is hermitian so that the
$\lambda_n$ are real.  Moreover, because $iD$ anticommutes with
$\gamma_5$ all eigenvalues occur in pairs $\pm\lambda$ so that $\rho$
is an even function.  The chiral condensate $\Xi$ is related to the
spectral density via the Banks-Casher formula~\cite{Bank80}
\begin{equation}
  \label{eq1.0}
  \Xi=\frac{\pi}{V}\rho(0) \:,
\end{equation}
where $V$ is the space-time volume.  The microscopic spectral density
is defined by~\cite{Shur93}
\begin{equation}
  \label{eq0.1}
  \rho_s(u)=\lim_{V\rightarrow\infty}\frac{1}{V\Xi} 
  \rho(\frac{u}{V\Xi}) \:.
\end{equation}
Based on sum rules derived by Leutwyler and Smilga~\cite{Leut92},
$\rho_s$ was conjectured to be a universal quantity~\cite{Shur93}.
This conjecture was supported by several
works~\cite{Verb93,Naga93,Verb96a,Jack96b,Slev93,Brez96,Akem96}.  As a
universal function, the microscopic spectral density can be computed
in the framework of random matrix models.  This task was performed at
zero temperature for all three chiral
ensembles~\cite{Verb93,Verb94b,Naga93}.  The most direct evidence for
the universality of $\rho_s$ came from a recent comparison of lattice
data obtained by Berbenni-Bitsch and Meyer with predictions from
random matrix theory where impressive agreement was
obtained~\cite{Wett97}.

Recently, the chiral random matrix models were extended to finite
temperature $T$ and finite chemical potential
$\mu$~\cite{Jack96a,Wett96,Step96a,Step96b,Nowa96}.  This leads to the
addition of a deterministic diagonal matrix to the random matrix used
at zero $T$ and $\mu$.  In such models, one can investigate, e.g., how
the chiral condensate decreases as a function of $T$ and/or $\mu$.
For the simplest model in which the additional diagonal matrix is
given by the lowest Matsubara frequency $\pi T$ and, thus,
proportional to the unit matrix, the spectral density has been
computed in Refs.~\cite{Step96a,Jack96b} and its microscopic limit in
Ref.~\cite{Jack96b}.  In the present paper, we (i) generalize these
results to arbitrary diagonal matrices and (ii) compute all higher
spectral correlation functions, both on the scale of the mean level
spacing and in the microscopic limit.  The first extension is
important since it is often desirable to consider more general
diagonal matrices, e.g., if one wants to include higher Matsubara
frequencies or in the approach of Ref.~\cite{Wett96} where the
formation of instanton--anti-instanton molecules was modeled.  The
second extension allows for further analysis of lattice results for,
e.g., scalar susceptibilities.  Our results also add further evidence
to the conjecture that the microscopic spectral correlations are
universal. 

This paper is organized as follows.  In Sec.~\ref{sec1} we define the
model and summarize some known results which will be refered to in the
following.  In Sec.~\ref{sec2} we set up the computation of the
spectral correlation functions in the framework of the graded
eigenvalue method and express the result in terms of a double
integral.  This double integral is evaluated in Sec.~\ref{sec3} for
the spectral density and the spectral correlations on the scale of the
mean level spacing and in Sec.~\ref{sec4} for the microscopic limit of
the spectral correlation functions, respectively.  The results are
discussed in Sec.~\ref{sec5}.  Three appendices are provided to
discuss convergence issues and technical details associated with
saddle-point approximations in Sec.~\ref{sec4}.

\section{Formulation of the problem}
\label{sec1}

We formulate the problem in the framework of random matrix theory
in Sec.~\ref{sec1a}. In Sec.~\ref{sec1b}, we discuss the connection
with Dyson's Brownian motion model.

\subsection{Definition of the model}
\label{sec1a}

We are interested in studying the spectral correlations of the Dirac
operator on the scale of the mean level spacing and in the microscopic
limit.  Since the spectral correlations are universal in these two
limits they can be obtained in the framework of a random matrix
model~\cite{Shur93}.  In such a model, the matrix of the Dirac
operator (in euclidean space) is replaced by a random matrix with
appropriate symmetries, the average over gauge field configurations is
replaced by the average over random matrices, and the complicated
weight function containing the euclidean action is replaced by a
simple Gaussian distribution of the random matrices.  Moreover, at
finite temperature a deterministic diagonal matrix with real entries
has to be added to the random matrix of the Dirac
operator~\cite{Jack96a,Wett96,Step96a,Nowa96}.  (Note that one could
also add a non-diagonal deterministic matrix since it could always be
diagonalized by a suitable rotation of the basis states.  This is
possible due to the invariance of the random matrix ensembles with
respect to basis changes.)

The matrix representing the Dirac operator in a chiral basis has the
structure
\begin{equation}
  \label{eq2.1}
  A=\left[\matrix{im\openone_N & W+Y \cr W^\dagger+Y &
  im\openone_N}\right] \:,
\end{equation}
where $m$ is an additional mass term, $W$ is a square random matrix of
dimension $N$, and $Y={\rm diag}(y_1,\ldots,y_N)$ is a diagonal matrix
with real entries.  The total number of eigenvalues of $A$ is $2N$
which can be identified with the space-time volume $V$ in
Eqs.~(\ref{eq1.0}) and (\ref{eq0.1}).  The specific form of $Y$ is
model dependent, but we will obtain results for an arbitrary choice of
$Y$.  Note that we consider the quenched case where $N_f$, the number
of flavors, is equal to zero.  Furthermore, we consider only square
matrices $W$.  Using rectangular matrices one can model the
topological charge $\nu$ which corresponds to the difference between
the number of rows and columns of $W$.  The extension of the present
calculation to non-zero $N_f$ and $\nu$ is in progress.

In this paper, we study the chiral GUE appropriate for QCD with three
or more colors for which $W$ is a complex matrix.  The probability
distribution of $W$ is a Gaussian,
\begin{equation}
  \label{eq2.2}
  P(W)=\left(\frac{N\Sigma^2}{\pi}\right)^{N^2}
       \exp\left(-N\Sigma^2{\rm tr}WW^\dagger\right) \:,
\end{equation}
where $\Sigma$ is a real parameter equal to the chiral condensate at
zero temperature.

We shall consider separately the case where all the $y_n$ are equal,
i.e., where $Y=y\openone_N$ (this will be refered to as the ``special
case'') and the case where $Y$ is arbitrary (this is the ``general
case'').  This separation is useful since the various mathematical
structures are most easily identified in the special case from which
they can then be generalized.

The chiral condensate $\Xi$, which depends on the matrix $Y$, has
already been computed for both the special case~\cite{Jack96a},
\begin{equation}
  \label{eq1.1}
  \Xi=\Sigma\sqrt{1-(\Sigma y)^2} \:,
\end{equation}
and the general case~\cite{Wett96},
\begin{equation}
  \label{eq1.2}
  \Xi=\Sigma^2\bar x \:.
\end{equation}
Here, $\bar x$ is the only real and positive solution of
\begin{equation}
  \label{eq1.3}
  \Sigma^2=\frac{1}{N}\sum_{n=1}^N\frac{1}{y_n^2+\bar x^2}
\end{equation}
or zero if no such solution exists.  (See Appendix~\ref{appB} for a
proof that this equation has at most one real and positive solution.)

The eigenvalue density and its microscopic limit for the special case
were computed in Ref.~\cite{Jack96b}.  In the following, we will
compute the eigenvalue density and the spectral correlation functions
on the scale of the mean level spacing as well as the corresponding
microscopic limits for the general case.

The $k$-point correlation functions measure the probability of finding
energies around the points $x_1,\ldots,x_k$, regardless of labeling.
Apart from trivial $\delta$-functions they are defined by
\begin{equation}
  \label{eq2.3}
  R_k(x_1,\ldots,x_k)=\left(-\frac{1}{\pi}\right)^k
  \int d[W] P(W) \prod_{p=1}^k {\rm Im \ tr} \frac{1}{x_p^+-A} \:, 
\end{equation}
where $x_p^+=x_p+i\varepsilon$ with $\varepsilon$ positive
infinitesimal.  In particular, we have $\rho(\lambda)=R_1(\lambda)$.
It is technically simpler to compute the functions
\begin{equation}
  \label{eq2.5}
  \hat R_k(x_1,\ldots,x_k)=\left(-\frac{1}{\pi}\right)^k
  \int d[W] P(W) \prod_{p=1}^k {\rm tr} \frac{1}{x_p^+-A} 
\end{equation}
from which the $R_k$ can be reconstructed by taking the imaginary
parts of the traces.

\subsection{Joint probability density function of the radial
  coordinates}
\label{sec1b}

Dyson's Brownian motion model of random matrix
theory~\cite{Meht91,Dyso62} describes the transition from an arbitrary
to a Gaussian ensemble. Since this approach is often used, we
establish the link to our random matrix model. The main modification
is of course due to the chiral structure.  In Ref.~\cite{Guhr96c}, we
gave a detailed discussion of a diffusion equation in the space of
arbitrary complex ordinary and supermatrices.  In the case of ordinary
matrices, this diffusion is the chiral analogue for the chGUE of
Dyson' Brownian motion for the GUE.

The pseudo-diagonalization of the matrix $W$ can be written as $W=UX
\bar V$, where $U\in{\rm U}(N)$, $\bar V\in{\rm U}(N)/ {\rm
  U}^{N}(1)$, and $X={\rm diag}(x_1,\ldots,x_N)$.  The $x_n$ are
referred to as radial coordinates, they are defined on the positive
real axis.  The transformation of the Cartesian volume element to
radial and angular coordinates reads~\cite{Guhr96c}
\begin{equation}
  \label{eq1b1}
  d[W] = \Delta_N^2(X^2)\prod_{n=1}^N x_n dx_n
                 d\mu(U)d\mu(\bar V) \:,
\end{equation}
where $\Delta_N(X^2)=\prod_{n<m}(x_n^2-x_m^2)$ is the Vandermonde
determinant and $d\mu(U)$ and $d\mu(\bar V)$ are the invariant Haar
measures.

Consider now $Y=0$ in Eqs.~(\ref{eq2.1}) and~(\ref{eq2.3}).
Obviously, the integrand depends only on the radial coordinates and
the group integrals are trivial. This so-called rotation invariance is
broken for non-zero $Y$. Nevertheless, even in this case, the
mathematical problem can be reduced to integrals over the radial
coordinates. By shifting the real part of the diagonal of $W$ and thus
also of $W^\dagger$ by the matrix $Y$, the breaking of the rotation
invariance is removed from the traces in Eq.~(\ref{eq2.3}). The
resulting probability density function $P(W-Y)$, however, now depends
explicitly on the angular coordinates $U$ and $\bar V$. To reduce the
problem to the radial coordinates, one has to perform the group
integrations
\begin{equation}
  \label{eq1b2}
  \Gamma(X,Y) = \int d\mu(U) \int d\mu(\bar V) P(UX\bar V-Y)
\end{equation}
which can be viewed as the extension of the Itzykson-Zuber integral to
complex matrices. In Refs.~\cite{Guhr96c,Jack96c},
this integral was calculated,
\begin{equation}
  \label{eq1b3}
\Gamma(X,Y) = \frac{1}{N!} 
             \frac{\det[\gamma(x_n,y_m)]_{n,m=1,\ldots,N}}
                  {\Delta_N(X^2)\Delta_N(Y^2)} \:,
\end{equation}
where the entries of the determinant are given by the functions
\begin{equation}
  \label{eq1b4}
\gamma(x_n,y_m) = 2N\Sigma^2 
           \exp\left(-N\Sigma^2(x_n^2+y_m^2)\right)
            I_0\left(2N\Sigma^2x_ny_m\right) \:.
\end{equation}
Here, $I_0(z)$ is the modified Bessel function of zeroth order. 

Collecting everything, the $k$-level correlation function is obtained
by integrating the distribution
\begin{equation}
  \label{eq1b5}
\Gamma(X,Y)\Delta_N^2(X^2)\prod_{n=1}^N x_n 
   =  \frac{1}{N!} \det[\gamma(x_n,y_m)]_{n,m=1,\ldots,N}
           \frac{\Delta_N(X^2)}{\Delta_N(Y^2)}\prod_{n=1}^N x_n 
\end{equation}
over the $N-k$ radial coordinates $x_{k+1},\ldots,x_N$. The
expression~(\ref{eq1b5}) is called the joint probability density
function. Thus, we have identified the kernel of the diffusion
equation of Ref.~\cite{Guhr96c} with the joint probability density
function of our random matrix model.  Apparently, it is highly
non-trivial to actually perform the integrations over the $N-k$ radial
coordinates.  In Sec.~\ref{sec2b}, we will show that the
supersymmetric analogue of the joint probability density function can
be used for a much faster and explicit calculation of the spectral
correlation functions.

\section{Correlation functions and supersymmetry}
\label{sec2}

After the introduction of random matrix theory by
Wigner~\cite{Wign51}, many of the mathematical details were worked out
by Dyson~\cite{Dyso62} and Mehta~\cite{Meht91}.  In 1983, Efetov
introduced the supersymmetry method~\cite{Efet83} which was further
developed in Ref.~\cite{Verb85}. In Sec.~\ref{sec2a} we employ these
methods to construct a supersymmetric representation for the spectral
correlation functions. It turns out that the formal aspects of the
problem discussed here are related to the transition from Poisson
regularity to the GUE. In Ref.~\cite{Guhr96a}, this
transition was discussed with supersymmetry by extending the graded
eigenvalue method developed in Ref.~\cite{Guhr91}. In Sec.~\ref{sec2b}
we further generalize this approach to the chGUE. The main ingredient
is now the supersymmetric extension of the Itzykson-Zuber like
integral~\cite{Guhr96c} mentioned in the previous section.

\subsection{Supersymmetric representation}
\label{sec2a}

The correlation functions~(\ref{eq2.3}) and~(\ref{eq2.5}) can be
obtained from the generating function
\begin{equation}
  \label{eq2.6}
  Z_k(J) = \int d[W] P(W) \frac{\prod_{p=1}^k\det(x_p^++J_p-A)}
  {\prod_{p=1}^k\det(x_p^+-J_p-A)} \:,
\end{equation}
where $J$ stands for $J_1,\ldots,J_k$.  Note that $Z_k(0)=1$.  The
$\hat R_k$ are then generated by
\begin{equation}
  \label{eq2.7}
  \hat R_k(x_1,\ldots,x_k)=\left(-\frac{1}{2\pi}\right)^k 
  \left.\frac{\partial^k}
  {\prod_{p=1}^k \partial J_p}Z_k(J) \right|_{J_p=0} \:.
\end{equation}
The ratio of the products of determinants in Eq.~(\ref{eq2.6}) can be
written as an inverse graded determinant which, in turn, can be
expressed as an integral over graded vectors.  We define
$N$-dimensional vectors $z_{p1}$, $z_{p2}$, $\zeta_{p1}$,
$\zeta_{p2}$, and $2N$-dimensional vectors $z_p=(z_{p1},z_{p2})^T$ and
$\zeta_p=(\zeta_{p1},\zeta_{p2})^T$.  The $z_p$ contain commuting
variables whereas the $\zeta_p$ contain anticommuting variables.
These variables are combined in a graded vector
$\psi=(z_1,\ldots,z_k,\zeta_1,\ldots,\zeta_k)^T$ of dimension
$4kN$.  The generating function (\ref{eq2.6}) can then be written as
\begin{eqnarray}
  \label{eq2.8}
  Z_k(J)&=&\int d[W] P(W)\ {\rm detg}^{-1} D \nonumber \\
  &=&\int d[W] P(W) \int d[\psi] \exp\left(i\psi^\dagger D\psi\right) \:,
\end{eqnarray}
where $D=(x^++J)\otimes\openone_{2N}-\openone_{2k}\otimes A$ is a
matrix of dimension $4kN$.  Here, $x+J={\rm diag}(x_1-J_1,\ldots,
x_k-J_k,x_1+J_1,\ldots,x_k+J_k)$ is a graded matrix of dimension $2k$,
and our convention for the direct product of two matrices $S$ and $T$
with number of rows $r_S$, $r_T$ and number of columns $c_S$, $c_T$ is
that \mbox{$(S\otimes T)_{(i-1)r_T+k,(j-1)c_T+l}=S_{ij}T_{kl}$}.  The
integration over $W$ can now be performed.  We write
\begin{equation}
  \label{eq2.9}
  \psi^\dagger \openone_{2k}\otimes\left[\matrix{0 & W \cr
    W^\dagger & 0} \right] \psi = {\rm tr}\left(WE+W^\dagger
  E^\dagger\right)
\end{equation}
with
\begin{equation}
  \label{eq2.10}
  E=\sum_{p=1}^k(z_{p2}z_{p1}^\dagger-\zeta_{p2}\zeta_{p1}^\dagger)
\end{equation}
and obtain the ensemble average
\begin{eqnarray}
  \int &d[W]&P(W) \exp\left(-i{\rm tr}(WE+W^\dagger
  E^\dagger)\right) \nonumber \\
  \label{eq2.11}
  && = \exp\left(-\frac{1}{N\Sigma^2}{\rm tr}
   EE^\dagger\right) = \exp\left(-\frac{1}{N\Sigma^2}{\rm
   trg} F_1F_2\right) \:.
\end{eqnarray}
The $F_i$ ($i=1,2$) are $2k$-dimensional graded matrices of the
structure
\begin{equation}
  \label{eq2.12}
  F_i=\left[\matrix{F_i^{BB}&F_i^{BF}\cr F_i^{FB}&F_i^{FF}}\right] \:,
\end{equation}
where the boson-boson and fermion-fermion block both have dimension
$k$.  The entries are given by scalar products
\begin{eqnarray}
  \label{eq2.13}
  (F_i^{BB})_{pp'}=z_{pi}^\dagger z_{p'i} & \phantom{WWW} &
  (F_i^{BF})_{pp'}=\zeta_{pi}^\dagger z_{p'i} \nonumber \\ 
  (F_i^{FB})_{pp'}=z_{pi}^\dagger \zeta_{p'i} &&
  (F_i^{FF})_{pp'}=\zeta_{pi}^\dagger \zeta_{p'i}
\end{eqnarray}
with $p,p'=1,\ldots,k$.  We now perform a Hubbard-Stratonovitch
transformation to decouple $F_1$ and $F_2$,
\begin{eqnarray}
  &\exp&\left(-\frac{1}{N\Sigma^2}{\rm trg} F_1F_2\right)\nonumber \\
  \label{eq2.14} 
  &&= 2^{2k^2}\int d[\sigma] \exp\left(-N\Sigma^2{\rm trg}\sigma
  \sigma^\dagger+i\,{\rm trg}(F_1\sigma+F_2\sigma^\dagger)\right) \:,
\end{eqnarray}
where $\sigma$ is a complex graded matrix of the same structure as
$F_i$.  Combining Eqs.~(\ref{eq2.8}), (\ref{eq2.11}), and
(\ref{eq2.14}), we obtain
\begin{equation}
  \label{eq2.15}
  Z_k(J)=2^{2k^2}\int d[\psi]\int d[\sigma]\exp\left(-N\Sigma^2{\rm trg}\sigma
  \sigma^\dagger+i\psi^\dagger G\psi\right)
\end{equation}
with
\begin{equation}
  \label{eq2.16} \!\!\!\!\!\!\!\!\!\!\! 
  G=\sigma\otimes\left[\matrix{\openone_N & 0\cr 0&0}\right]
    +\sigma^\dagger\otimes\left[\matrix{0&0\cr 0&\openone_N}\right] 
    +(x^+\!\!+\!J)\otimes\openone_{2N}-\openone_{2k}\otimes\left[
    \matrix{0&Y\cr Y&0}\right] \:.
\end{equation}
Note that we have set the quark mass $m$ equal to zero in
(\ref{eq2.16}).  It can always be reinstated at the end of the
calculation if we shift the $x_p$ by $im$.  

We would now like to interchange the order of the $\sigma$- and the
$\psi$-integrations in (\ref{eq2.15}) and to perform the integration
over $\psi$.  This would be allowed if the $\psi$-integral were
uniformly convergent.  Uniform convergence is guaranteed for the
ordinary GUE but not for the chGUE.  In fact, the $\psi$-integral as
it stands is not uniformly convergent.  This presents certain problems
which are discussed in some detail in Appendix~\ref{app0}.  There are
various ways to proceed.  For the present calculation, we note that
there is a very strong conjecture that ignoring the convergence
problem will lead to a result whose real part is incorrect but whose
imaginary part is correct.  Since we are only interested in the
spectral correlation functions which are given by the imaginary part,
we choose to go ahead with the calculation without modifications.
However, the reader should consult Appendix~\ref{app0} for a more
complete discussion of this problem.

Ignoring the convergence issue, we now shift $\sigma$ and
$\sigma^\dagger$ by $-(x+J)$ and perform the integration over $\psi$
to obtain after some algebra
\begin{eqnarray}
  \label{eq2.17}
  Z_k(J)=2^{2k^2}\int d[\sigma]&&\exp\left(-N\Sigma^2{\rm trg}
  (\sigma-(x+J))(\sigma^\dagger-(x+J))\right) \nonumber \\
  &&\times \ {\rm detg}^{-1}\left(\sigma^+(\sigma^+)^\dagger
  \otimes\openone_N- \openone_{2k}\otimes Y^2\right) \:.
\end{eqnarray}
In order to work out the integrals over the supermatrix, there are 
various ways to proceed. In particular, Efetov's~\cite{Efet83} coset
method has been used in numerous applications. It is based on a 
saddle-point approximation in the large-$N$ limit. To solve the
problems outlined in the introduction we use the graded eigenvalue
method~\cite{Guhr91,Guhr96a} which is, in this context, 
particularly well suited.

\subsection{Reduction to eigenvalue integrals}
\label{sec2b}

The matrix $\sigma$ can be written in pseudo-diagonal form as
$\sigma=us\bar v$, where $u\in{\rm U}(k/k)$, $\bar v\in{\rm U}(k/k)/
{\rm U}^{2k}(1)$, and $s={\rm diag}(s_1,is_2)$ with $s_j={\rm
  diag}(s_{1j},\ldots, s_{kj})$ for $j=1,2$~\cite{Guhr96c}.  The
$s_{pj}$ are real and non-negative.  The transformation of the
Cartesian volume element to radial and angular coordinates
reads~\cite{Guhr96c}
\begin{eqnarray}
  \label{eq2.18}
  d[\sigma]&=&J(s)d[s]d\mu(u)d\mu(\bar v) \nonumber \\
  d[s]&=&\prod_{p=1}^k ds_{p1} ds_{p2} \nonumber \\
  J(s)&=&B_{k}^2(s^2) \prod_{p=1}^k s_{p1}s_{p2}
\end{eqnarray}
with
\begin{eqnarray}
  \label{eq2.19}
  B_{k}(s)&=&\frac{\Delta_{k}(s_1)\Delta_{k}(is_2)}
  {\prod_{p,q}(s_{p1}-is_{q2})} \\
  \Delta_k(x)&=&\prod_{p<q}(x_p-x_q) \:.
\end{eqnarray}
The group measure is the invariant Haar measure.  We obtain
\begin{eqnarray}
  \label{eq2.20} \!\!\!\!\!\!
  Z_k(J)&=&2^{2k^2}\exp\left(-N\Sigma^2{\rm trg}(x+J)^2\right) 
  \nonumber \\
  &&\times\int d[s] J(s)\exp(-N\Sigma^2{\rm trg}s^2) {\rm detg}^{-1} 
  \left((s^+)^2\otimes\openone_N-\openone_{2k}\otimes Y^2\right) 
  \nonumber \\
  && \times\int d\mu(u) \int d\mu(\bar v)
  \exp\left(N\Sigma^2{\rm trg}(\sigma+\sigma^\dagger)(x+J)\right) \:.
\end{eqnarray}
The integral over the diagonalizing groups has been computed
in Ref.~\cite{Guhr96c}.  The result is
\begin{eqnarray}
  \label{eq2.21} \!\!\!\!\!\!
  \int &&d\mu(u) \int d\mu(\bar v) \exp\left(N\Sigma^2{\rm trg}
  (\sigma+\sigma^\dagger)(x+J)\right) \nonumber \\ 
  &&=\frac{(2N\Sigma^2)^{2k}}{2^{2k^2}(k!)^2}
  \frac{\det[I_0(2N\Sigma^2s_{p1}(x_{p'}\!-\!J_{p'}))]
  \det[J_0(2N\Sigma^2s_{q2}(x_{q'}\!+\!J_{q'}))]}
  {B_{k}(s^2)B_{k}((x+J)^2)}
\end{eqnarray}
with $p,p'=1,\ldots,k$ and $q,q'=1,\ldots,k$.  Here, $J_0$ and $I_0$
are the Bessel and modified Bessel functions of order 0, respectively.
Thus, Eq.~(\ref{eq2.20}) becomes
\begin{eqnarray}
  \label{eq2.22} 
  Z_k(J)&=&1+\frac{(2N\Sigma^2)^{2k}\exp\left(-N\Sigma^2{\rm trg}
  (x\!+\!J)^2\right)}{(k!)^2B_{k}((x\!+\!J)^2)}\int d[s]\prod_{p=1}^k 
   s_{p1}s_{p2} B_{k}(s^2) \nonumber \\
  && \times\exp(-N\Sigma^2{\rm trg}s^2) {\rm detg}^{-1}\left((s^+)^2
  \otimes\openone_N-\openone_{2k}\otimes Y^2\right) \nonumber \\
  && \times \det[I_0(2N\Sigma^2s_{p1}(x_{p'}-J_{p'}))]_{p,p'=
   1,\ldots,k} \nonumber \\
  && \times \det[J_0(2N\Sigma^2s_{q2}(x_{q'}+J_{q'}))]_{q,q'=
   1,\ldots,k} \:. 
\end{eqnarray}
Note the contribution of unity to the generating function.  It ensures
the correct normalization of $Z_k$ at $J=0$ and is due to so-called
Efetov-Wegner terms which were discussed in detail in
Ref.~\cite{Guhr96c}.

The calculation now greatly simplifies due to the determinant
structure of the function $B_k(s)$.  In fact, Eq.~(\ref{eq2.19}) can
be rewritten as
\begin{equation}
  \label{eq2.23}
  B_k(s)=\det\left[\frac{1}{s_{p1}-is_{q2}}\right]_{p,q=1,\ldots,k} \:.
\end{equation}
Performing the differentiations with respect to the $J_p$ according to
Eq.~(\ref{eq2.7}) and using the determinant structure of $Z_k$,
straightforward manipulations show that the $k$-point functions can be
written as
\begin{equation}
  \label{eq2.24}
  R_k(x_1,\ldots,x_k)=\det[C_N(x_p,x_q)]_{p,q=1,\ldots,k} \:,
\end{equation}
where the function $C_N$ is given by
\begin{eqnarray}
  \label{eq2.25}
  C_N(x_p,x_q)&=&\frac{2}{\pi}(2N\Sigma^2)^2x_p
  \int_0^\infty\int_0^\infty ds_1ds_2 \frac{s_1s_2}{s_1^2+s_2^2}
  \exp\left(-N\Sigma^2(s_1^2+s_2^2)\right) \nonumber \\
  &&\times I_0(2N\Sigma^2x_ps_1)J_0(2N\Sigma^2x_qs_2) \
  {\rm Im}\prod_{n=1}^N\frac{y_n^2+s_2^2}{y_n^2-(s_1+i\varepsilon)^2} \:.
\end{eqnarray}
Hence, all $k$-point functions are known if the double integral
(\ref{eq2.25}) can be computed.  This can be done by saddle-point
approximation in the large-$N$ limit.  However, for some applications
it is more convenient to rewrite the double integral as an integral
over a $2\times 2$ complex graded matrix $\sigma$ by using our result
(\ref{eq2.21}) for the integral over the diagonalizing groups in the
opposite direction.  This is by no means a trivial step since we needed
the angular integral in order to obtain the determinant structure of
(\ref{eq2.24}).  Using Eq.~(\ref{eq2.21}) for $k=1$, we obtain from
(\ref{eq2.25})
\begin{eqnarray}
  \label{eq2.26a} \!\!\!\!\!\!\!\!\!\!\!\!
  &&C_N(x_p,x_q)=\frac{8}{\pi}x_p
  \frac{e^{N\Sigma^2(x_p^2-x_q^2)}}{x_p^2-x_q^2} \
  {\rm Im} \int d[\sigma] \exp\left(-N{\cal L}(\sigma,\sigma^\dagger)
  \right) \\ 
  \label{eq2.26b} \!\!\!\!\!\!\!\!\!\!\!\!
  &&{\cal L}(\sigma,\sigma^\dagger)=\Sigma^2{\rm trg}\sigma
  \sigma^\dagger - \frac{1}{N}{\rm trg\ log}\left((\sigma^+\!+\!x)
  ((\sigma^+)^\dagger\!+\!x)\otimes\openone_N
  -\openone_2\otimes Y^2\right) \:,
\end{eqnarray}
where $\sigma$ is now a $2\times 2$ complex graded matrix and $x={\rm
  diag}(x_p,x_q)$.  Note that $\sigma$ and $\sigma^\dagger$ have been
shifted by $x$ to obtain Eqs.~(\ref{eq2.26a}) and (\ref{eq2.26b}).
The integral in Eq.~(\ref{eq2.26a}) can also be computed in
saddle-point approximation in the large-$N$ limit.  Since the matrix
$\sigma$ can be pseudo-diagonalized, $\sigma=us\bar v$, we can assume
that $\sigma$ is diagonal at the saddle point, $\bar\sigma={\rm
  diag}(s_1,is_2)$, and thus obtain from (\ref{eq2.26b})
\begin{eqnarray}
  \label{eq2.28}
  {\cal L}(s_1,is_2)=\Sigma^2(s_1^2+s_2^2) - \frac{1}{N}\sum_{n=1}^N
  \log\frac{(is_2+x_q)^2-y_n^2}{(s_1+x_p)^2-y_n^2} \:,
\end{eqnarray}
where we have omitted the imaginary increments.

We will use Eq.~(\ref{eq2.26a}) to compute the spectral density and
the universal spectral correlations in the bulk of the spectrum in
Sec.~\ref{sec3} and Eq.~(\ref{eq2.25}) to compute the microscopic
limits of the spectral density and the spectral correlation functions
in Sec.~\ref{sec4}.  Our expressions allow for a relatively
straightforward finite-$N$ analysis.  However, we are eventually
interested in the thermodynamic limit.  Therefore, we compute our
results in the limit $N\rightarrow\infty$.

\section{Spectral density and universal spectral correlations in the
  bulk of the spectrum}
\label{sec3}

In this section, we compute the spectral density and the universal
spectral correlations in the bulk of the spectrum on the scale of the
mean level spacing, for both the special and the general case.  The
mean level spacing $D(x)$ is defined as $D(x)=1/\rho(x)$.
With our definitions, we have $D(x)\sim 1/(N\Sigma)$ in the bulk of
the spectrum.  Thus, the correlations are to be computed for energy
differences $\sim 1/(N\Sigma)$. We discuss the special case in
Sec.~\ref{sec3.2} and turn to the general case in Sec.~\ref{sec3.3}.

\subsection{Special case}
\label{sec3.2}

In the special case $Y=y\openone_N$ with $y$ real, (\ref{eq2.28})
becomes 
\begin{equation}
  \label{eq3.2.1}
  {\cal L}(s_1,is_2)=\Sigma^2(s_1^2+s_2^2)-\log\frac{(is_2+x_q)^2-y^2}
  {(s_1+x_p)^2-y^2} \:.
\end{equation}
To minimize $\cal L$, we differentiate, first with respect to $s_1$,
\begin{equation}
  \label{eq3.2.2}
  \frac{\partial {\cal L}}{\partial s_1}=2\Sigma^2s_1+
  \frac{2(s_1+x_p)}{(s_1+x_p)^2-y^2}\stackrel{!}{=} 0 
\end{equation}
which yields a cubic equation for $s_1$.  Differentiation of $\cal L$
with respect to $is_2$ yields an identical equation with $s_1$ and
$x_p$ replaced by $is_2$ and $x_q$, respectively.  To compute the
spectral density, we have to take the limit $x_p=x_q$.  To compute the
universal spectral correlations, $x_p-x_q$ is of order $1/(N\Sigma)$
and, thus, $x_p=x_q$ in the large-$N$ limit.  Hence, the saddle-point
equations yield identical solutions for $s_1$ and $is_2$, and we
define $s_1=is_2\equiv\bar s$ at the saddle point.  Furthermore, we
define $x=(x_p+x_q)/2$ and $\Delta x=x_p-x_q$.  Note that $x_p=x_q=x$
for $N\rightarrow\infty$.  The saddle-point equation then reads
\begin{equation}
  \label{eq3.2.3}
  {\bar s}^3+2{\bar s}^2x+{\bar s}(1/\Sigma^2+x^2-y^2)+x/\Sigma^2=0 \:.
\end{equation}
The Lagrangian (\ref{eq3.2.1}) at the saddle point can then be
expanded in $\Delta x$,
\begin{equation}
  \label{eq3.2.4}
  {\cal L}_0=-\log\frac{(\bar s+x-\Delta x/2)^2-y^2}
  {(\bar s+x+\Delta x/2)^2-y^2} 
  \approx \frac{2(\bar s+x)}{(\bar s+x)-y^2}\Delta x 
  = -2\Sigma^2\bar s\Delta x \:,
\end{equation}
where we have used Eq.~(\ref{eq3.2.2}) in the last step.  The integral
over the quadratic fluctuations yields a factor of $1/4$ (cf.~our
convention for the normalization of the Gaussian in
Eq.~(\ref{eq2.14})), and we obtain from Eq.~(\ref{eq2.26a})
\begin{eqnarray}
  \label{eq3.2.5}
  C_N(x_p,x_q)&=&\frac{8}{\pi}x_p
  \frac{e^{N\Sigma^2(x_p^2-x_q^2)}}{x_p^2-x_q^2}
  \,\frac{1}{4}\,{\rm Im}\,\exp\left(-N{\cal L}_0\right) \nonumber \\
  &\approx&\frac{e^{2N\Sigma^2x\Delta x}}
  {\pi\Delta x}\,{\rm Im}\,\exp(2N\Sigma^2\bar s\Delta x)\nonumber\\
  &=&\frac{\sin(2N\Sigma^2\Delta x{\rm Im}\,\bar s)}
  {\pi\Delta x}\exp(2N\Sigma^2(x+{\rm Re}\,\bar s)\Delta x) \:.
\end{eqnarray}
Because of the identity 
\begin{equation}
  \label{eq3.2.6}
  \det\left[f(x_p,x_q)\exp\left(\alpha(x_p-x_q)\right)
  \right]_{p,q=1,\ldots,k}=\det\left[f(x_p,x_q)\right]_{p,q=1,\ldots,k}
\end{equation}
for arbitrary $f$ and $\alpha$ we obtain in the large-$N$ limit
\begin{equation}
  \label{eq3.2.7}
  R_k(x)=\det[K_N(x_p,x_q)]_{p,q=1,\ldots,k} 
\end{equation}
with
\begin{equation}
  \label{eq3.2.8}
  K_N(x_p,x_q)=\frac{\sin(2N\Sigma^2(x_p-x_q){\rm Im}\,\bar s)}
  {\pi(x_p-x_q)} \:.
\end{equation}
The spectral density now follows immediately,
\begin{equation}
  \rho(x)=K_N(x,x)=\frac{2N\Sigma^2}{\pi}\,{\rm Im}\,\bar s \:.
  \label{eq3.2.9}
\end{equation}
The cubic equation (\ref{eq3.2.3}) always has one real solution.  If
the other two solutions are real, $K_N(x_p,x_q)$ and $\rho(x)$ are
zero for this value of $x$.  Otherwise, the other two solutions are
complex conjugates of each other, and we have to take the solution
with positive imaginary part because of the positive imaginary
increment of $s$.  This is the solution of (\ref{eq3.2.3}) which
enters (\ref{eq3.2.8}) and (\ref{eq3.2.9}). 

Furthermore, we obtain from (\ref{eq3.2.9}) 
\begin{equation}
  \label{eq3.2.10}
  D(x)=\frac{\pi}{2N\Sigma^2{\rm Im}\,\bar s}
\end{equation}
so that (\ref{eq3.2.8}) can be rewritten as
\begin{equation}
  \label{eq3.2.11}
  D(x)K_N(x_p,x_q)=\frac{\sin\left(\pi(x_p-x_q)/D(x)\right)}
  {\pi(x_p-x_q)/D(x)} \:.
\end{equation}
Eq.~(\ref{eq3.2.11}) reproduces the well-known result for the spectral
correlations on the scale of the mean level spacing.  Thus, neither
the chiral structure of our random matrix model nor the addition of
the diagonal matrix $Y$ changes these universal correlations.

We would like to emphasize that the result for $\rho(\lambda)$ has
been obtained previously in a simpler
way~\cite{Step96a,Jack96b,Nowa96}.  However, our method has the virtue
of also yielding all higher-order correlations and their microscopic
limits in one single formalism.

\subsection{General case}
\label{sec3.3}

Consider now the case of general (but real) $Y$ where the Lagrangian
has the form (\ref{eq2.28}).  Again, we minimize ${\cal L}$ and obtain
identical solutions for $s_1$ and $is_2$ in the limit $x_p=x_q$.  With
the definitions of the previous subsection, the equation for $\bar s$
is now
\begin{equation}
  \label{eq3.3.1}
  \Sigma^2\bar s+\frac{1}{N}\sum_{n=1}^N\frac{\bar s+x}
  {(\bar s+x)^2-y_n^2}=0 \:.
\end{equation}
This is a polynomial equation of order $2N+1$.  We now demonstrate
that this equation always has at least $2N-1$ real solutions.  The
remaining two solutions are either real or complex conjugates of each
other. 

Define $z=\bar s+x$ and consider the function
\begin{equation}
  \label{eq3.3.2}
  f(z)=\Sigma^2(z-x)+\frac{1}{N}\sum_{n=1}^N\frac{z}{z^2-y_n^2} \:.
\end{equation}
First assume that all the $y_n$ are different and non-zero.  The
function $f(z)$ has $2N$ poles at $\pm y_n$ ($n=1,\ldots,N$).  Now
consider two adjacent poles $z_m<z_{m+1}$.  We have
$f(z_m+\varepsilon)>0$ and $f(z_{m+1}-\varepsilon)<0$, where
$\varepsilon$ is positive infinitesimal.  Since $f(z)$ is continuous
for $z_m<z<z_{m+1}$, there must be a zero of $f$ in this interval.
There are $2N-1$ such intervals, hence there are $2N-1$ real zeros of
$f$.  This proves our assertion.  If some of the $y_n$ are equal or
zero, the saddle-point equation (\ref{eq3.3.1}) is of correspondingly
lower order, and our argument applies as well.

In analogy to the result of the previous subsection, only the (single)
saddle point with positive imaginary part contributes to
$C_N(x_p,x_q)$.  Expanding the Lagrangian about this saddle point
yields
\begin{eqnarray}
  {\cal L}_0&=&-\frac{1}{N}\sum_{n=1}^N\log\frac{(\bar s+x
  -\Delta x/2)^2-y_n^2}{(\bar s+x+\Delta x/2)^2-y_n^2} 
  \approx \frac{2}{N}\sum_{n=1}^N
  \frac{(\bar s+x)}{(\bar s+x)-y_n^2}\Delta x \nonumber \\
  &=& -2\Sigma^2\bar s\Delta x 
  \label{eq3.3.3}
\end{eqnarray}
using (\ref{eq3.3.1}) in the last step.  Going through the same steps
as in the previous subsection, we obtain the same results as before,
the only difference being that $s$ is now the only solution with
positive imaginary part (or zero) of the more complicated equation
(\ref{eq3.3.1}) instead of the cubic equation (\ref{eq3.2.3}).

\section{Microscopic limit}
\label{sec4}

In this section, we compute the $k$-point functions in the microscopic
limit, i.e., for $x_p\sim 1/(N\Sigma)$.  In Sec.~\ref{sec4.0}, we
separate the radial integrals in (\ref{eq2.25}), then consider the
special case $Y=y\openone_N$ in Sec.~\ref{sec4.1}, and finally turn to
the general case in Sec.~\ref{sec4.2}.  Although the special case
obviously follows from the general case, we find it useful to discuss
the special case first in order to make the structure of the radial
integrals more explicit.  The final result will be that the $k$-point
functions are given by the zero-temperature result provided that the
microscopic variables are rescaled by the temperature-dependent chiral
condensate.

\subsection{Separation of the radial integrals}
\label{sec4.0}

We first define microscopic variables $u_p=2Nx_p$ and $u_q=2Nx_q$.
The two integrals in Eq.~(\ref{eq2.25}) can be decoupled by writing
\begin{equation}
  \label{eq4.0.1}
  \frac{\exp\left(-N\Sigma^2(s_1^2+s_2^2)\right)}{s_1^2+s_2^2}=
  N\Sigma^2\int_1^\infty dt\exp\left(-N\Sigma^2t(s_1^2+s_2^2)\right) \:.
\end{equation}
We thus obtain
\begin{equation}
  \label{eq4.0.2}
  C_N(u_p,u_q)=\frac{4N^2\Sigma^6}{\pi}u_p\int_1^\infty dtf_1(t)f_2(t)
\end{equation}
with
\begin{eqnarray}
  \label{eq4.0.3a}
  f_1(t)&=&\int_0^\infty ds_1s_1\exp\left(-N\Sigma^2ts_1^2\right)
  I_0(\Sigma^2u_ps_1)\,{\rm Im}\prod_{n=1}^N
  \frac{1}{y_n^2-(s_1+i\varepsilon)^2} \\
  \label{eq4.0.3b}
  f_2(t)&=&\int_0^\infty ds_2s_2\exp\left(-N\Sigma^2ts_2^2\right)
  J_0(\Sigma^2u_qs_2)\prod_{n=1}^N(y_n^2+s_2^2) \:.
\end{eqnarray}
These two integrals can now be computed in saddle-point approximation.

\subsection{Special case}
\label{sec4.1}

Consider the special case $Y=y\openone_N$ with $y$ real.  The critical
value of $y$ above which the chiral condensate vanishes is given by
$y_c=1/\Sigma$ so that we restrict ourselves to the interval $0\le y
\le y_c$ in the following.  Eqs.~(\ref{eq4.0.3a}) and (\ref{eq4.0.3b})
become 
\begin{eqnarray}
  \label{eq4.1.1a}
  f_1(t)&=&\int_0^\infty ds_1s_1\exp\left(-N\Sigma^2ts_1^2\right)
  I_0(\Sigma^2u_ps_1)\,{\rm Im}
  \frac{1}{(y^2-(s_1+i\varepsilon)^2)^N} \\
  \label{eq4.1.1b}
  f_2(t)&=&\int_0^\infty ds_2s_2\exp\left(-N\Sigma^2ts_2^2\right)
  J_0(\Sigma^2u_qs_2)(y^2+s_2^2)^N  \:.
\end{eqnarray}
These two integrals can be evaluated in saddle-point approximation in
the large-$N$ limit.  This somewhat technical task is performed in
Appendix~\ref{appA}.  It turns out that depending on the value of $t$,
the product of $f_1(t)$ and $f_2(t)$ can be of different order in $1/N$.  
Specifically, there is a ``critical'' value of $t$,
\begin{equation}
  \label{eq4.1.2}
  t_c=\frac{1}{(\Sigma y)^2} \:,
\end{equation}
below which $f_1(t)f_2(t)$ is of order $1/N$ and above which
$f_1(t)f_2(t)$ is suppressed exponentially.  Note that $t_c\ge 1$
since $y\le y_c$.  Thus, only the interval $1\le t\le t_c$ contributes
to the integral in Eq.~(\ref{eq4.0.2}) in leading order in $1/N$.  We
obtain from Eqs.~(\ref{eqA.6}) and (\ref{eqA.11})
\begin{equation}
  \label{eq4.1.5} 
  f_1(t)f_2(t)=\frac{\pi}{4N\Sigma^4t^2}
  J_0\left(\Sigma u_p\sqrt{1/t-1/t_c}\right)
  J_0\left(\Sigma u_q\sqrt{1/t-1/t_c}\right)
\end{equation}
to leading order in $1/N$ in this region of $t$.  Substituting
$z=\Sigma\sqrt{1/t-1/t_c}$ and using
$\Xi=\Sigma\sqrt{1-1/t_c}$, we have from Eq.~(\ref{eq4.0.2})
\begin{eqnarray}
  \label{eq4.1.6}
  C_N(u_p,u_q)&=&2Nu_p\int_0^\Xi dz z J_0(u_pz)J_0(u_qz) \\
  \label{eq4.1.7}
  &=& 2N\Xi \tilde u_p\frac{\tilde u_pJ_1(\tilde u_p)J_0(\tilde u_q)-
            \tilde u_qJ_0(\tilde u_p)J_1(\tilde u_q)}
           {\tilde u_p^2-\tilde u_q^2} \:,
\end{eqnarray}
where we have defined $\tilde u_p=\Xi u_p$ and $\tilde u_q=\Xi u_q$
and used Eq.~(11.3.29) of Ref.~\cite{Abra72}.  In particular, we
obtain for $k=1,2$
\begin{eqnarray}
  \label{eq4.1.8}
  \rho_s(\tilde u)&=&\frac{\tilde u}{2}\left(J_0^2(\tilde u)
    +J_1^2(\tilde u)\right) \\
  \label{eq4.1.9}
  \rho_s(\tilde u_1,\tilde u_2)&=&\tilde u_1\tilde u_2 
    \left(\frac{\tilde u_1J_1(\tilde u_1)J_0(\tilde u_2)-
    \tilde u_2J_0(\tilde u_1)J_1(\tilde u_2)}
    {\tilde u_1^2-\tilde u_2^2}\right)^2 \:,
\end{eqnarray}
where we have subtracted the disconnected part of the two-point
function in (\ref{eq4.1.9}).  Eq.~(\ref{eq4.1.8}) is in agreement with
the result of Ref.~\cite{Jack96b}.  The zero-temperature limit of
Eq.~(\ref{eq4.1.9}) agrees with the result obtained in
Ref.~\cite{Verb93}.

\subsection{General case}
\label{sec4.2}

We now generalize the results of the previous subsection to the case
where $Y={\rm diag}(y_1,\ldots,y_N)$.  We only require that the $y_n$
be real.  The chiral condensate for this case has been computed in
Ref.~\cite{Wett96}, and the condition for the existence of a
condensate was found to be
\begin{equation}
  \label{eq4.2.1}
  \Sigma^2\le\frac{1}{N}\sum_{n=1}^N \frac{1}{y_n^2} \:.
\end{equation}
Note that there is always a condensate if some of the $y_n$ are zero
(or more precisely, if a finite fraction of the $y_n$ is zero).  We
shall restrict ourselves to choices of $Y$ for which the above
condition is satisfied.

We are now dealing directly with the two integrals in
Eqs.~(\ref{eq4.0.3a}) and (\ref{eq4.0.3b}).  In Appendix~\ref{appB},
these two integrals are evaluated in saddle-point approximation in the
large-$N$ limit. Again, $f_1(t)f_2(t)$ is of order $1/N$ for $t<t_c$
and suppressed exponentially for $t>t_c$, respectively, where now
\begin{equation}
  \label{eq4.2.3}
  t_c=\frac{1}{\Sigma^2}\frac{1}{N}\sum_{n=1}^N \frac{1}{y_n^2} \:.
\end{equation}
Note that $t_c\ge 1$ by (\ref{eq4.2.1}) and that $t_c$ is infinite if
some of the $y_n$ are zero.  Again, only the interval $1\le t\le t_c$
contributes to leading order in $1/N$, and we obtain from
Eqs.~(\ref{eqA.6}) and (\ref{eqA.11}) using Eqs.~(\ref{eqB.7a}) and
(\ref{eqB.7b})
\begin{equation}
  \label{eq4.2.4}
  f_1(t)f_2(t)=\frac{\pi}{N} J_0(\Sigma^2u_p\bar x)
  J_0(\Sigma^2u_q\bar x)\frac{\bar x^2}{{\cal L}_2''(\bar x)}
\end{equation}
to leading order in $1/N$ in this region of $t$, where ${\cal
  L}_2''(\bar x)$ is given by Eq.~(\ref{eqB.5b}) and $\bar x$ is
implicitly given by Eq.~(\ref{eq4.2.6}) below.  To proceed, we
substitute $z=\Sigma^2\bar x$ and obtain
\begin{equation}
  \label{eq4.2.5}
  dz=\Sigma^2\frac{d\bar x}{dt}dt \:.
\end{equation}
To compute $d\bar x/dt$, we use the saddle-point equation
(\ref{eqB.2}) in the form
\begin{equation}
  \label{eq4.2.6}
  \Sigma^2t=\frac{1}{N}\sum_{n=1}^N\frac{1}{y_n^2+\bar x^2} 
\end{equation}
and differentiate with respect to $t$ to obtain
\begin{equation}
  \label{eq4.2.7}
  \Sigma^2=-2\bar x\frac{1}{N}\sum_{n=1}^N\frac{1}{(y_n^2+\bar x^2)^2}
  \frac{d\bar x}{dt}=-\frac{1}{2\bar x}{\cal L}_2''(\bar x)
  \frac{d\bar x}{dt} \:,
\end{equation}
where we have used Eq.~(\ref{eqB.5b}) in the last equality.  Simple
algebra yields
\begin{equation}
  \label{eq4.2.8}
  \frac{\bar x^2}{{\cal L}_2''(\bar x)}dt=-\frac{1}{2\Sigma^6}zdz
\end{equation}
so that Eq.~(\ref{eq4.0.2}) in connection with Eq.~(\ref{eq4.2.4})
reduces to Eq.~(\ref{eq4.1.6}) with $\Xi=\Sigma^2\bar x(t=1)$, where
$\bar x(t=1)$ is the solution of (\ref{eq4.2.6}) for $t=1$.  We have
thus reduced the general case to the special case considered in the
previous subsection.  The only complication in the general case is
that one has to solve the higher-order polynomial equation
(\ref{eq4.2.6}) for $\bar x(t=1)$.  This is no problem numerically
since we have shown in Appendix~\ref{appB} that there is only one real
and positive solution which has to be determined.

\section{Discussion}
\label{sec5}

We have considered a random matrix model appropriate for the Dirac
operator of QCD at finite temperature and derived the spectral
density, the universal spectral correlations on the scale of the mean
level spacing, and the microscopic limits of the spectral density and
spectral correlation functions.  The calculation was done for an
arbitrary real diagonal matrix added to the random matrix of the Dirac
operator.  The fact that the functional form of the microscopic
correlations remains unchanged even with an arbitrary deterministic
matrix added adds further evidence to the conjecture that these
correlations (and in particular the microscopic spectral density) are
universal. 

It is a well known feature of standard random matrix theory that many
of the conceptual and formal aspects are common to all three
ensembles.  In particular, the calculation of fluctuations shows these
common features since the scale is always set by the mean level
spacing.  Therefore, we conjecture that the universal scaling behavior
which we derived for all correlations in the chGUE carries over to the
chGOE and the chGSE.  More precisely, we expect that, after proper
rescaling with the chiral condensate, the microscopic spectral
correlations are given by the known zero-temperature results for the
chGOE and the chGSE, respectively~\cite{Verb94b,Naga93}.

There are two important directions in which one would like to
generalize the results of this paper.  First, one would like to extend
our results to a non-zero number of flavors and non-zero topological
charge.  This presents certain technical challenges to the graded
eigenvalue method, but we are convinced that an application of the
Itzykson-Zuber integral for rectangular supermatrices~\cite{Guhr96c}
will solve this problem.  The result, however, can be conjectured now:
After proper rescaling with the chiral condensate, we will obtain the
same dependence on $N_f$ and $\nu$ as in the zero-temperature result
computed in Ref.~\cite{Verb93}.  Second, one would like to consider
complex diagonal matrices instead of just real ones in order to model
the effect of the chemical potential on the spectrum of the Dirac
operator.  For the latter case, it is important to go to $N_f\ne 0$,
i.e., to the unquenched case, as was argued by
Stephanov~\cite{Step96b}.  At present, there is quite some effort in
this direction~\cite{Jani97}.

Note: After completion of the present work, we learned that
a similar calculation was performed simultaneously and independently
by Jackson, \c Sener, and Verbaarschot.

\section*{Acknowledgments}

We would like to thank A.\ M\"uller-Groeling, A.\ Sch\"afer,
J.J.M.\ Verbaarschot, and H.A.\ Weidenm\"uller for useful
discussions.

\appendix

\section{Uniform convergence}
\label{app0}

The integral over $\psi$ in Eq.~(\ref{eq2.15}) does not converge
uniformly since the matrix $\sigma$ is arbitrary complex rather than
hermitian so that its eigenvalues are complex rather than real.  Note
that it is sufficient to consider only the commuting variables of
$\sigma$ since the anticommuting variables cannot cause convergence
problems.  Hence, the $\sigma$- and the $\psi$-integration in
(\ref{eq2.15}) cannot simply be interchanged.

The obvious way to proceed is to deform the integration contour of the
$\sigma$-variables so that uniform convergence is obtained.  This was
done in Ref.~\cite{Jack96b} where the one-point function was computed
by an explicit re-parameterization of the $\sigma$-variables.  This
entails replacing compact integration variables by non-compact ones.
The generalization of this approach to higher correlation functions is
not trivial.  One would have to pseudo-diagonalize $\sigma$ using
non-compact groups and compute the resulting integrals of the
Itzykson-Zuber type. 

However, there is an easier way to proceed.  In the calculation of the
one-point function, one observes that a naive interchange of the
$\sigma$- and $\psi$-integrations, while upsetting the real part of
the result, leaves the imaginary part of the result
intact~\cite{Jack96b}.  The imaginary part of the result is all that
is needed to construct the one-point function.  Hence, as far as the
one-point function is concerned, we {\em can} interchange the
$\sigma$- and $\psi$-integrations and still get the correct result.
While there is no proof at present that this remarkable feature
persists for higher correlation functions, we conjecture that it does.
This conjecture is very strongly supported by the fact that we obtain
the correct zero-temperature result for the higher correlation
functions by simply interchanging the order of the integrations as we
have done in the text.  Since the convergence problem is completely
unrelated to the introduction of the matrix $Y$ at finite
temperature, it is hard to see how the presence of $Y$ could change
anything in this respect.

Of course, this problem deserves further attention.  Work in this
direction is in progress, but there is no doubt that the results
presented in this paper will be unaffected by a clarification of this
issue.

\section{Saddle-point approximation for the special case}
\label{appA}

We wish to compute the two integrals in Eqs.~(\ref{eq4.1.1a}) and
(\ref{eq4.1.1b}) in the large-$N$ limit by saddle-point approximation.
An interesting phenomenon occurs since the integrals depend on an
external parameter $t$.  We shall see that the integrals can be of
different order in $1/N$ depending on the value of $t$.

Let us start with the integral in Eq.~(\ref{eq4.1.1b}) since it is
conceptually simpler.  Combining terms which grow exponentially with
$N$, we obtain
\begin{eqnarray}
  \label{eqA.1a}
  f_2(t)&=&\int_0^\infty ds_2 s_2J_0(\Sigma^2u_qs_2)
  \exp\left(-N {\cal L}_2(s_2)\right) \\
  \label{eqA.1b}
  {\cal L}_2(s_2)&=&\Sigma^2ts_2^2-\log(y^2+s_2^2) \:.
\end{eqnarray}
To minimize ${\cal L}_2(s_2)$, we differentiate
\begin{equation}
  \label{eqA.2}
  \frac{\partial {\cal L}_2}{\partial s_2}=2\Sigma^2
  ts_2-\frac{2s_2}{y^2+s_2^2}\stackrel{!}{=} 0 \:.
\end{equation}
There are three possible solutions for $s_2$, $\bar s_2=0$ and $\bar
s_2=\pm\bar x$ with
\begin{equation}
  \label{eqA.3}
  \bar x=\frac{1}{\Sigma} \sqrt{\frac{1}{t}-\frac{1}{t_c}} \:,
\end{equation}
where $t_c=1/(\Sigma y)^2$.  Expanding about the saddle points yields
\begin{equation}
  \label{eqA.4}
  {\cal L}_2(\bar s_2+\delta s_2)={\cal L}_2(\bar s_2)
   +\frac{1}{2}{\cal L}_2''(\bar s_2) (\delta s_2)^2
\end{equation}
with 
\begin{eqnarray}
  \label{eqA.5a}
  {\cal L}_2(0)=&-\log(y^2) \phantom{2\Sigma^2(t-t_c)} &
  {\cal L}_2(\pm\bar x)=1-t/t_c+\log(\Sigma^2t) \\
  \label{eqA.5b}
  {\cal L}_2''(0)=&2\Sigma^2(t-t_c) \phantom{-\log(y^2)} &
  {\cal L}_2''(\pm\bar x)=4\Sigma^2t(1-t/t_c) \:.
\end{eqnarray}
The sign of ${\cal L}_2''$ determines the location of the minimum of
${\cal L}_2$.  For $t>t_c$, the minimum is at $0$.  For $t<t_c$, the
minimum is at $\bar x$, and $\bar x$ is real.  The solution $-\bar x$
lies outside the interval of integration and can be discarded.  We now
have to integrate over the fluctuations.  For $\bar s_2=\bar x$, we
obtain to leading order in $1/N$
\begin{eqnarray}
  \label{eqA.6}
  f_2(t)&=&\bar x J_0(\Sigma^2u_q\bar x) e^{-N{\cal L}_2(\bar x)}
  \int_{-\infty}^\infty d(\delta s_2) e^{-\frac{N}{2}{\cal L}_2''(\bar
  x)(\delta s_2)^2} \nonumber \\
  &=& \sqrt{\frac{2\pi}{N}}e^{-N{\cal L}_2(\bar x)}
  J_0(\Sigma^2u_q\bar x) \frac{\bar x}{\sqrt{{\cal L}_2''(\bar x)}}
  \qquad {\rm for} \quad t<t_c \:.
\end{eqnarray}
For $\bar s_2=0$, we obtain to leading order in $1/N$
\begin{eqnarray}
  \label{eqA.7}
  f_2(t)&=&e^{-N{\cal L}_2(0)} \int_0^\infty d(\delta s_2) \delta s_2
  e^{-\frac{N}{2}{\cal L}_2''(0)(\delta s_2)^2}  \nonumber \\
  &=&\frac{1}{N}\frac{e^{-N{\cal L}_2(0)}}{{\cal L}_2''(0)}
  \qquad {\rm for} \quad t>t_c \:.
\end{eqnarray}

Now consider the integral in Eq.~(\ref{eq4.1.1a}).  We again combine
terms which grow exponentially with $N$ to obtain
\begin{eqnarray}
  \label{eqA.8a}
  f_1(t)&=&{\rm Im} \int_0^\infty ds_1 s_1I_0(\Sigma^2u_ps_1)
  \exp\left(-N {\cal L}_1(s_1)\right) \\
  \label{eqA.8b}
  {\cal L}_1(s_1)&=&\Sigma^2ts_1^2+\log(y^2-s_1^2) \:,
\end{eqnarray}
where we have omitted the imaginary increment of $s_1$ in ${\cal
  L}_1$.  Differentiation yields
\begin{equation}
  \label{eqA.9}
  \frac{\partial {\cal L}_1}{\partial s_1}=2\Sigma^2
  ts_1-\frac{2s_1}{y^2-s_1^2}\stackrel{!}{=} 0 \:.
\end{equation}
This equation is identical to Eq.~(\ref{eqA.2}) if $s_1$ is replaced
by $is_2$.  Hence, possible solutions are $\bar s_1=0$ and $\bar
s_1=\pm i\bar x$.  Expansion about the saddle points yields
\begin{eqnarray}
  \label{eqA.10a}
  {\cal L}_1(0)=&\log(y^2) \phantom{2\Sigma^2(t-t_c)} &
  {\cal L}_1(\pm i\bar x)=t/t_c-1-\log(\Sigma^2t) \\
  \label{eqA.10b}
  {\cal L}_1''(0)=&2\Sigma^2(t-t_c) \phantom{\log(y^2)} &
  {\cal L}_1''(\pm i\bar x)=4\Sigma^2t(1-t/t_c) \:.
\end{eqnarray}
Note that ${\cal L}_1(i\bar s_2)=-{\cal L}_2(\bar s_2)$.  Although the
saddle-point equations (\ref{eqA.2}) and (\ref{eqA.9}) look remarkably
similar, the structure of the saddle points is now quite different.
In the following, we show that the saddle point at $i\bar x$ ($-i\bar
x$) makes the dominant contribution for $t<t_c$ ($t>t_c$).

Consider first the case $t<t_c$ where $\bar x>0$.  The possible saddle
points $\pm i\bar x$ are purely imaginary.  Hence, we have to deform
the contour of integration into the complex plane to reach these
saddle points.  Due to the positive sign of the imaginary increment of
$s_1$, only the point $+i\bar x$ can be reached.  Since ${\cal
  L}_1''(i\bar x)$ is positive for $t<t_c$, we have to integrate over
quadratic fluctuations parallel to the real axis.  Since the original
range of integration was 0 to $\infty$ and since the saddle point lies
on the imaginary axis, the integration over the fluctuations is only
from 0 to $\infty$.  We obtain for the contribution from the saddle
point $i\bar x$
\begin{eqnarray}
  \label{eqA.11}
  f_1(t)&=&{\rm Im}\ i\bar x I_0(\Sigma^2u_pi\bar x) 
  e^{-N{\cal L}_1(i\bar x)} \int_0^\infty d(\delta s_1) 
  e^{-\frac{N}{2}{\cal L}_1''(i\bar x)(\delta s_1)^2} \nonumber \\
  &=& \sqrt{\frac{\pi}{2N}}e^{-N{\cal L}_1(i\bar x)}
  J_0(\Sigma^2u_p\bar x) \frac{\bar x}{\sqrt{{\cal L}_1''(i\bar x)}}
  \qquad {\rm for} \quad t<t_c \:.
\end{eqnarray}
The saddle point $\bar s_1=0$ does not contribute at all since the
imaginary part of the product of the value of the integrand at the
saddle point and the integral over the fluctuations is zero in this
case.

Now consider the case $t>t_c$ where $\bar x$ is imaginary.  The
relevant saddle point is now $\bar s_1=-i\bar x$ which is real and
positive.  The point $i\bar x$ is real and negative and, hence,
outside the interval of integration.  Since ${\cal L}_1''(-i\bar x)$
is negative for $t>t_c$, we now have to integrate over quadratic
fluctuations parallel to the imaginary axis from 0 to $\infty$. We
obtain
\begin{eqnarray}
  \label{eqA.12}
  f_1(t)&=&{\rm Im}\ (-i\bar x) I_0(-\Sigma^2u_pi\bar x) 
  e^{-N{\cal L}_1(-i\bar x)} \int_0^\infty d(i\delta s_1) 
  e^{-\frac{N}{2}{\cal L}_1''(-i\bar x)(i\delta s_1)^2} \nonumber \\
  &=& \sqrt{\frac{\pi}{2N}}e^{-N{\cal L}_1(-i\bar x)}
  J_0(\Sigma^2u_p\bar x) \frac{-i\bar x}{\sqrt{-{\cal L}_1''(-i\bar x)}}
  \qquad {\rm for} \quad t>t_c
\end{eqnarray}
which is identical in structure with Eq.~(\ref{eqA.11}).  Again, the
saddle point $\bar s_1=0$ does not contribute for the same reason as
in the case $t<t_c$.

Combining Eqs.~(\ref{eqA.6}) and (\ref{eqA.11}) we obtain
Eq.~(\ref{eq4.1.5}) which is valid for $t<t_c$.  Hence,
$f_1(t)f_2(t)\sim 1/N$ in this region of $t$.  For $t>t_c$,
Eqs.~(\ref{eqA.7}) and (\ref{eqA.12}) combine to yield
\begin{equation}
  \label{eqA.14}
  f_1(t)f_2(t)=\frac{\sqrt{\pi}}{\sqrt{32}N^{3/2}\Sigma^4tt_c}
  I_0\left(\Sigma u_p\sqrt{1/t_c-1/t}\right)
  \frac{e^{-N\left(\frac{t}{t_c}-1-\log\frac{t}{t_c}\right)}}{t/t_c-1} 
\end{equation}
showing that $f_1(t)f_2(t)$ is suppressed exponentially in this case.

We would like to add that we have checked Eqs.~(\ref{eqA.11}) and
(\ref{eqA.12}) as well as the corresponding results of
Appendix~\ref{appB} by an independent finite-$N$ calculation and
taking the limit $N\rightarrow\infty$.

\section{Saddle-point approximation for the general case}
\label{appB}

We now compute the more general integrals in Eqs.~(\ref{eq4.0.3a}) and
(\ref{eq4.0.3b}) in the large-$N$ limit by saddle-point approximation.
We shall encounter the same phenomena as in Appendix~\ref{appA}.  For
simplicity, we restrict ourselves to the case where all $y_n$ are
different and non-zero as in Sec.\ref{sec3.3}.  It is straightforward
to modify our arguments to work for special cases (some of the $y_n$
are equal or zero) as well.

Again, we write Eq.~(\ref{eq4.0.3b}) in the form (\ref{eqA.1b}), where
now
\begin{equation}
  \label{eqB.1}
  {\cal L}_2(s_2)=\Sigma^2ts_2^2-\frac{1}{N}\sum_{n=1}^N
  \log(y_n^2+s_2^2) \:.
\end{equation}
To minimize ${\cal L}_2(s_2)$, we differentiate
\begin{equation}
  \label{eqB.2}
  \frac{\partial {\cal L}_2}{\partial s_2}=2\Sigma^2 ts_2-2s_2
  \frac{1}{N}\sum_{n=1}^N\frac{1}{y_n^2+s_2^2}\stackrel{!}{=} 0 \:.
\end{equation}
The obvious solution $\bar s_2=0$ persists.  There are now many more
solutions for $s_2^2$ ($N$ of them, to be precise), but we show in the
following that there is only one real and positive solution for
$s_2^2$ if $t<t_c$ and no real and positive solution for $s_2^2$ if
$t>t_c$, where $t_c$ is given by Eq.~(\ref{eq4.2.3}).  Define
\begin{equation}
  \label{eqB.3}
  g(x)=t-\frac{1}{\Sigma^2N}\sum_{n=1}^N\frac{1}{y_n^2+x} \:.
\end{equation}
We seek real and positive solutions of $g(x)=0$.  We have
\begin{eqnarray}
  \label{eqB.4}
  g'(x)&=&\frac{1}{\Sigma^2N}\sum_{n=1}^N\frac{1}{(y_n^2+x)^2}>0 
  \nonumber \\
  \lim_{x\rightarrow\infty}g(x)&=&t>0 \nonumber \\
  g(0)&=&t-t_c\left\{\begin{array}{l}
                      <0 \quad {\rm for} \quad t<t_c \\
                      >0 \quad {\rm for} \quad t>t_c \:.
                   \end{array}\right.
\end{eqnarray}
The function $g(x)$ increases monotonically for $x>0$ and is
continuous in this region.  It reaches a positive value as
$x\rightarrow\infty$.  Thus, $g(x)$ intersects the $x$-axis once and
only once if $g(0)<0$ and not at all if $g(0)>0$.  This proves our
claim.

Let $\bar x$ be the only real and positive solution of $g(x^2)=0$ for
$t<t_c$.  Thus, $\bar s_2=\bar x$ is the only other candidate for a
minimum of ${\cal L}_2$.  Expansion about the saddle points as in
Eq.~(\ref{eqA.4}) yields
\begin{eqnarray}
  \label{eqB.5a}
  {\cal L}_2(0)=&-\frac{1}{N}\sum_{n=1}^N\log(y_n^2) 
    \phantom{2\Sigma^2(t-t_c)} \!\!\!\!\!\!\!\!\!\!\!\! &
  {\cal L}_2(\bar x)=\Sigma^2t\bar x^2-\frac{1}{N}\sum_{n=1}^N
    \log(y_n^2+\bar x^2) \\
  \label{eqB.5b}
  {\cal L}_2''(0)=&2\Sigma^2(t-t_c) \phantom{-\frac{1}{N}\sum_{n=1}^N
    \log(y_n^2)} \!\!\!\!\!\!\!\!\!\!\!\! &
  {\cal L}_2''(\bar x)=4\bar x^2\frac{1}{N}\sum_{n=1}^N\frac{1}
    {(y_n^2+\bar x^2)^2} \:.
\end{eqnarray}
The explicit form of ${\cal L}_2(0)$ and ${\cal L}_2(\bar x)$ will not
be needed in the following.  We again obtain that the minimum of
${\cal L}_2$ is at $\bar x$  for $t<t_c$ and at 0 for $t>t_c$.
Integration over the fluctuations yields Eqs.~(\ref{eqA.6}) and
(\ref{eqA.7}) as before.

Writing Eq.~(\ref{eq4.0.3a}) in the form (\ref{eqA.8b}) yields
\begin{equation}
  \label{eqB.6}
  {\cal L}_1(s_1)=\Sigma^2ts_1^2+\frac{1}{N}\sum_{n=1}^N
  \log(y_n^2-s_1^2) \:.
\end{equation}
Note that ${\cal L}_1(is_2)=-{\cal L}_2(s_2)$.  For the
$s_1$-integration, there are now many more candidates for saddle
points.  Apart from the trivial solution $\bar s_1=0$ (which does not
contribute for the reason discussed in Appendix~\ref{appA}), there are
$2N$ other solutions for $\bar s_1$.  Defining a function similar to
$g(x)$ and investigating its pole structure, we find that for $t<t_c$
there are $N-1$ real and positive solutions for $\bar s_1^2$.  None of
these corresponds to a minimum of ${\cal L}_1$ on the real axis since
${\cal L}_1''<0$ at these points.  The remaining two solutions are
$\bar s_1=\pm i\bar x$, and expansion about the saddle points yields
\begin{eqnarray}
  \label{eqB.7a}
  {\cal L}_1(\pm i\bar x)&=&-{\cal L}_2(\bar x) \\
  \label{eqB.7b}
  {\cal L}_1''(\pm i\bar x)&=&{\cal L}_2''(\bar x) \:.
\end{eqnarray}
As in Appendix~\ref{appA}, the relevant saddle point is $\bar
s_1=i\bar x$, and integration over the fluctuations yields
Eqs.~(\ref{eqA.11}) as before.  Eqs.~(\ref{eqA.6}) and (\ref{eqA.11})
combine to yield Eq.~(\ref{eq4.2.4}).

For $t>t_c$, there are $N$ real solutions for $\bar s_1^2$ (apart from
$\bar s_1=0$).  The negative solution for $\bar s_1^2$ for $t<t_c$ has
now becomes positive, and it is straightforward to see from the
structure of the function $g(x)$ that it is now the smallest of all
possible solutions for $\bar s_1^2$.  Note that we still have ${\cal
  L}_1''<0$ at all possible saddle points.  We now deform the
integration contour as in Appendix~\ref{appA}: We move on the real
axis from 0 to the smallest positive saddle point and then up parallel
to the imaginary axis.  Only the smallest positive saddle point
contributes to $f_1(t)$, and we obtain a result similar to
Eq.~(\ref{eqA.12}), with $-i\bar x$ replaced by the smallest positive
saddle point.  The product $f_1(t)f_2(t)$ is again suppressed
exponentially for $t>t_c$ so that only the interval $1\le t\le t_c$
contributes to the integration over $t$ in Eq.~(\ref{eq4.0.2}).

\end{document}